\documentclass[twocolumn,aps,preprintnumbers,showpacs]{revtex4}
\usepackage{epsfig}

\begin{document}

\preprint{CU-TP-1171, KIAS-P06062, hep-th/yymmnnn}

\title{Hawking-Moss bounces and vacuum decay rates}
\author{Erick J. Weinberg}
\email{ejw@phys.columbia.edu}
 \affiliation{Physics Department, Columbia University, New York, New
York 10027, USA}
\affiliation{School of Physics, Korea Institute for Advanced Study,
207-43, Cheongnyangni-2dong, Dongdaemun-gu, Seoul 130-722,
Korea}


\begin{abstract}

The conventional interpretation of the Hawking-Moss (HM) solution implies a
transition rate between vacua that depends only on the values of the
potential in the initial vacuum and at the top of a potential barrier,
leading to the implausible conclusion that transitions to distant
vacua can be as likely as those to a nearby one.  I analyze this
issue using a nongravitational example with analogous
properties.  I show that such HM bounces do not give reliable rate
calculations, but are instead related to
the probability of finding a quasistable configuration at a
local potential maximum.

\end{abstract}

\maketitle

A quantum field theory may have, in addition to the ``true vacuum'' of
minimum energy, one or more metastable ``false vacua'' of higher
energy.  The latter can decay by nucleating, through either quantum
tunneling or thermal fluctuation, bubbles of true vacuum that then
expand and coalesce.  In a series of papers
\cite{Coleman:1977py,Callan:1977pt,Coleman:1980aw}, Coleman and
collaborators developed a formalism, based on ``bounce'' solutions of
the Euclideanized field equations, for calculating the rate of such
bubble nucleation.  When this formalism is extended to take gravity
into account, one finds not only the Coleman-De Luccia (CDL) bounces
\cite{Coleman:1980aw} that seem analogous to the flat-space bounces,
but also the spatially homogeneous Hawking-Moss (HM) bounce
\cite{Hawking:1981fz} that appears to represent a fluctuation from the false
vacuum to the top of the potential barrier separating the true and
false vacua.

A striking feature of the HM bounce is that, by the conventional
interpretation, it implies a transition rate that depends only on the
values of the potential at the top of the barrier and at the false
vacuum, but not at intermediate points.  For a theory with only a
single false vacuum, this may perhaps be seen as just a curiosity,
although it is certainly a bit troubling.
However, it becomes particularly salient when viewed in the context of
a possible string theory landscape.  If the transition rate depends
only on the initial and final values of the potential, then
transitions to distant parts of the landscape could be as likely as
those to nearby points.  Although the probability of any given
transition might be exponentially small, the exponentially large
number of potential final states could make the lifetime of any one
vacuum, even one with a small cosmological constant, quite short.
In this letter, I will re-examine the reasoning that leads to this 
conclusion, and show where it fails.

\begin{figure}[t]
\begin{center}
\epsfig{file=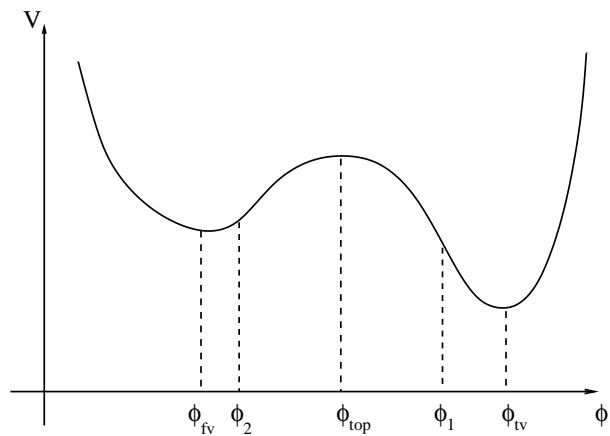, width =8cm }
\end{center}
\caption{The potential for a typical theory with a false
  vacuum. \label{firstfig}}
\end{figure}

To begin, let us recall the main features of the bounce formalism.
Consider the theory of a single scalar field $\phi$ with a potential
such as that shown in Fig.~\ref{firstfig}.  In the
absence of gravity, and at zero temperature, the bounce is a solution
of the four-dimensional Euclidean field equations with a region of
approximate true vacuum surrounded by an infinite false vacuum region.
For the case of an O(4)-symmetric bounce that depends on a single
radial variable $s$, we have $\phi(\infty) = \phi_{\rm fv}$, while
$\phi(0)$ lies on the true vacuum side of the potential barrier.  Note
that $\phi(0) \ne \phi_{\rm tv}$.  The rate per unit volume at which
true vacuum bubbles nucleate within a false vacuum region is of the
form $\Gamma  = A e^{-B}$,
where $B = S_E(\phi_{\rm bounce}) - S_E(\phi_{\rm fv})$ is the difference 
between the Euclidean action of the bounce solution and that of the homogeneous
configuration with $\phi = \phi_{\rm fv}$ everywhere.  (We will not need the
detailed expression for $A$ in this discussion.)

Gravity can be incorporated into this formalism by including the
Euclidean action for gravity in $S_E$, and solving for both the scalar
field and the metric.  If $V(\phi)$ is everywhere non-negative, then
the bounce solution necessarily has the topology of a four-sphere.  In
the CDL bounce the scalar field at one pole has a value that is on the
true vacuum side of the barrier, while at the antipodal point it is on
the false vacuum side.  In Fig.~\ref{firstfig}, $\phi_1$ and $\phi_2$ indicate
these values for a ``typical'' CDL bounce.   In limits where one 
would expect gravitational effects on bubble nucleation to be small, $|\phi_2
-\phi_{\rm fv}|$ is exponentially small and the CDL result for $B$ approaches
that of the flat space calculation.

In the HM solution the metric is the standard round metric
on a four-sphere of radius $H_{\rm top}^{-1} \equiv 
[8\pi V(\phi_{\rm top})/3 M_{\rm Pl}^2]^{-1/2}$, with
$\phi=\phi_{\rm top}$ everywhere on the four-sphere.  
A calculation of its Euclidean action leads to a tunneling exponent
\begin{equation}
    B_{\rm HM}= S_E(\phi_{\rm top}) - S_E(\phi_{\rm fv})  
       = -{3 M^4_{\rm Pl} \over 8V(\phi_{\rm top}) } 
               + {3 M^4_{\rm Pl} \over 8V(\phi_{\rm fv}) } \, .
\end{equation}
In the case where $[V(\phi_{\rm top}) - V(\phi_{\rm fv})]/V(\phi_{\rm fv})
\ll 1$, this can be approximated by 
\begin{equation}
    B_{\rm HM} \approx {[(4\pi/3) H_{\rm fv}^{-3}][V(\phi_{\rm top}) - V(\phi_{\rm fv})]
           \over T_{\rm fv} } \, ,
\label{thermalB}
\end{equation}
where $H_{\rm fv}^{-1} \equiv [8\pi V(\phi_{\rm fv})/3 M_{\rm
Pl}^2]^{-1/2}$ and $T_{\rm fv} = H_{\rm fv}/2\pi$ is the de Sitter
temperature of the false vacuum.  The form of the latter expression
suggests an interpretation of the HM bounce as corresponding to a
thermal fluctuation of a horizon-sized region up to the top of the
barrier, following which the field can roll (either homogeneously or
spinodally), down to either the false or the true vacuum
\cite{volumefoot}.  [There is no contribution from gravitational
energy in Eq.~(\ref{thermalB}) because it is obtained by taking the
limit in which the difference in $V$, and hence in the geometry,
between the false vacuum and the top of the barrier is small.]

\begin{figure}[b]
\begin{center}
\epsfig{file=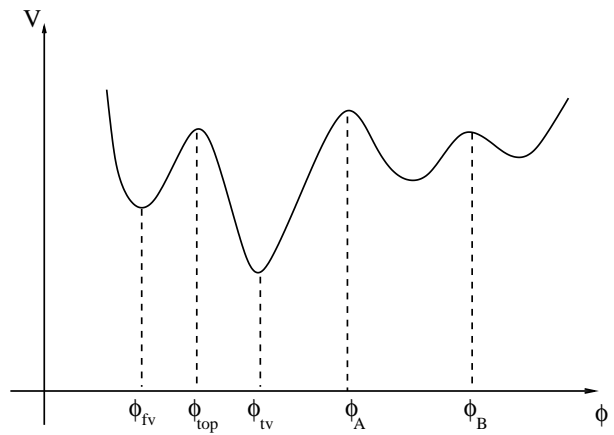, width =8cm }
\end{center}
\caption{A potential with several local maxima. \label{secondfig}}
\end{figure}

The difficulty, alluded to above, is that for a potential with many
local maxima there are HM bounces corresponding to each of these
maxima.  Thus, for the potential shown in Fig.~\ref{secondfig}, there
are HM solutions associated with the local maxima of $V$ at $\phi_A$
and $\phi_B$, in addition to the one associated with $\phi_{\rm top}$.
If any of these maxima are degenerate with the one at $\phi_{\rm
top}$, then the standard interpretation would imply that the
transitions from $\phi_{\rm fv}$ to these more distant points would
occur at the same rate as transitions to $\phi_{\rm top}$.

Since this conclusion seems intuitively untenable, one is led to ask
what might be wrong with the reasoning involved.  The first thought
might be that the HM bounce is somehow pathological, and that one
should ignore it and use only CDL bounces.  The difficulty with this
is that it is always possible to continuously deform the potential in
such a way that the endpoints of the CDL bounce move up toward, and
eventually meet at, $\phi_{\rm top}$ \cite{Jensen:1983ac,
Hackworth:2004xb}.  The CDL bounces with endpoints near $\phi_{\rm
top}$ will have counterparts that are concentrated near the maxima at
$\phi_A$ and $\phi_B$, and these will lead to conclusions similar to,
and as discomforting as, those following from the HM bounces.

At this point one might decide that gravity, especially in its
Euclidean form, is too subtle and that one should not trust the CDL
formalism; further support for this idea might come from noting that
this formalism was obtained by analogy with the flat space case rather
than through a rigorous derivation.  However, one is faced with the
fact that the CDL bounces seem to give reasonable results in the limit
where gravitational effects are small.

Furthermore, there is a non-gravitational example that also suffers
from the same difficulties.  Consider the scalar field theory
described above on a compact space that is a three-sphere of radius
$L$, with a nonzero temperature $T$.  In the finite temperature
extension of the bounce formalism \cite{Langer:1969bc,Linde:1981zj},
one looks for Euclidean solutions with periodicity $1/T$ in the
imaginary time.  For $T \gg L^{-1}$, the dominant bounces are constant
in imaginary time.  This effectively reduces the problem to a
three-dimensional one that will have homogeneous solutions completely
analogous to the four-dimensional HM bounces and, for suitable
potentials, CDL-type solutions that are confined to the region near
the maxima of the potential \cite{firstfoot}.  Thus, it is clear that
the difficulties we have found cannot be attributed to Euclidean
gravity.

These remarks lead us to ask the following questions:

1) Does the bounce formalism indeed go wrong for some or all of the HM, and nearby CDL,
bounces and, if so, why?

2) If a HM bounce doesn't calculate a transition rate, then what does 
it calculate?

In order to emphasize that the origin of these difficulties does not
lie in the gravitational setting, as well as to avoid some of the
conceptual issues associated with quantum field theory in curved
spacetime, I will address these questions using the finite volume,
finite temperature example (although I will adopt the HM and CDL
terminology to refer to the corresponding bounces).  

To begin, let us recall the reasoning that led to the bounce
formalism.  At zero temperature and infinite volume, with gravity
ignored, bubble nucleation is a quantum tunneling process in which one
tunnels through the potential energy barrier that separates a
spatially homogenous configuration with $\phi({\bf x}) = \phi_{\rm
fv}$ from a configuration that has a bubble of approximate true vacuum
surrounded by false vacuum exterior.  The one-dimensional WKB
treatment of quantum tunneling is readily extended to systems with
many degrees of freedom \cite{Banks:1973ps,Banks:1974ij}.  For each
possible path through the barrier one can define a WKB tunneling
exponent $B$.  The tunneling rate is obtained from the path that
minimizes $B$.  
By arguments familiar (apart from some changes of sign) from classical
mechanics, one can show \cite{Coleman:1977py} that this minimization
problem is equivalent to finding a stationary point of the Euclidean
action.  This stationary point is the bounce solution.

A more precise formulation is obtained by a path integral argument
\cite{Callan:1977pt}.  Before describing this, it will be best to
specify more clearly what is meant by the false vacuum.  This is
unambiguous classically, where it is the homogeneous state with
$\phi({\bf x}) = \phi_{\rm fv}$.  In the quantum field theory,
however, things are not so simple.  It is not sufficient to require
that the expection value of the field $\langle \phi({\bf x}) \rangle=
\phi_{\rm fv}$, because this condition is also satisfied by states
corresponding to mixed regions of approximate true and approximate
false vacua.  In addition, we want to require that the wave functional
be concentrated on configurations that approximate the spatially
homogeneous classical false vacuum.  This can be made more precise by
considering a deformed potential that agrees with $V(\phi)$ in the
potential well containing $\phi_{\rm fv}$ but that increases
monotonically outside this interval.  The theory defined by this
deformed potential has a stable ground state with the properties that
we require of the false vacuum state; I will denote this state by
$|{\rm FV}\rangle$.  More generally, the states $| n_{\rm fv} \rangle$
that are energy eigenstates of the deformed problem with energies small 
relative to barrier height
will have wave
functionals concentrated in the region surrounding $\phi_{\rm fv}$
and can be viewed as excited states built upon the false
vacuum.

Now consider the path integral
\begin{equation}
    I({\cal T}) = \int [d\phi] e^{-S_{\rm E}[\phi]}  = \langle \phi_{\rm fv}|
      e^{-H{\cal T}} |\phi_{\rm fv} \rangle   \, ,
\label{pathint}
\end{equation}
where the integration is over configurations satisfying $\phi({\bf
x},\pm {\cal T}/2) = \phi_{\rm fv}$, while $|\phi_{\rm fv}\rangle$
denotes the state whose wave functional is a delta functional at
$\phi({\bf x}) = \phi_{\rm fv}$.  The standard procedure would now be
to expand the matrix element on the right-hand side in terms of a
complete set of energy eigenstates and to note that in the limit
${\cal T}\rightarrow \infty$ the sum would be dominated by the term
corresponding to the lowest energy state.  This would be the true
vacuum, whereas we are interested in the false vacuum.  Note, however,
that the overlap of the true vacuum with $|\phi_{\rm fv}\rangle$ is
exponentially small.  Indeed, there is a similar suppression for any
state that is orthogonal to (or that has exponentially small inner 
product with) the $| n_{\rm fv} \rangle$ defined above.  Hence, for a
large range of $\cal T$ the sum is dominated not by the absolutely
lowest lying state, but rather by the lowest of the $| n_{\rm fv}
\rangle$ (i.e., by the false vacuum $|{\rm FV}\rangle$), giving
\begin{equation}
    I({\cal T}) \approx |\langle {\rm FV}|\phi_{\rm fv} \rangle|^2
    e^{-E_{\rm fv}{\cal T}} \, .
\end{equation}
After evaluating $I({\cal T})$ by expanding about its stationary
points (the classical false vacuum, the bounce, and the multibounce
solutions), one can read off $E_{\rm fv}$ from its large $\cal T$
behavior.  Because $| {\rm FV} \rangle$ (like all of the $| n_{fv}
\rangle$) is not quite an eigenstate of the full Hamiltonian, $E_{\rm
fv}$ is complex.  Its imaginary part corresponds to a decay rate
and yields the desired nucleation rate $\Gamma$.

The finite temperature tunneling rate is obtained not from the
imaginary part of the energy of the false vacuum, but rather from the
imaginary part of the free energy of the false vacuum
\cite{Linde:1981zj}.  This can be gotten by replacing the path
integral in Eq.~(\ref{pathint}) by a similar integral in which the
integration is over configurations that are periodic with period $1/T$
in imaginary time.  This gives the partition function $Z= {\rm Tr}\,
e^{-H/T}$, and thus the free energy \cite{secondfoot}.
An unrestricted
integral over all periodic configurations would correspond to a trace
over all states, and would give the full partition function for the
theory.  Since what we actually want is the partition function
corresponding to the metastable false vacuum, the trace must be
restricted to the subspace spanned by the $| n_{\rm fv} \rangle$.
Hence, the integration should only be over periodic configurations
with nontrivial overlap with the $| n_{\rm fv} \rangle$.

In the infinite volume case, this restriction is automatically imposed
by the requirement that the three-dimensional configurations have
finite energy, relative to that of the false vacuum, which implies
that they must approach $\phi_{\rm fv}$ at spatial infinity.  When the
volume is finite, on the other hand, any smooth configuration will
have finite energy.  Hence, a naive application of the bounce
formalism could well admit configurations that do not contribute to
$Z_{\rm fv}$, and that are therefore not relevant for a calculation of
the lifetime of the false vacuum.

We now see how the bounce formalism can go astray for a HM bounce.  In
order that the false vacuum be metastable (i.e., that it persist for a
substantial length of time), homogeneous configurations lying outside
the potential well surrounding $\phi_{\rm fv}$ must be far out in the
exponential tails of the wave functionals both of $|{\rm FV}\rangle$
and of the excited false vacuum states $|n_{\rm fv}\rangle$.  Hence, a
homogeneous three-dimensional configuration with $\phi$ at the top of
a distant local maxima is necessarily among those that must be
excluded, and so the corresponding HM bounce cannot lead to a reliable
calculation of $\Gamma$.

The situation is more subtle for the HM bounces corresponding to the
maxima that bound the false vacuum potential well.  These, too,
correspond to configurations with exponentially small overlap with
$|{\rm FV}\rangle$.  However, they can have nontrivial overlap with
those excited false vacuum $|n_{\rm fv}\rangle$ states with energies
comparable to the top of the barrier.  At finite temperature these
excited states will be populated, although that population will be
thermally suppressed.  In fact, for states with energy equal to the
top of the barrier, this suppression factor is (up to possible
pre-exponential factors) precisely $e^{-B_{\rm HM}}$.  Thus, one can
view this HM bounce as representing an evaporation process rather
than a tunneling one~\cite{Batra:2006rz}.  Alternatively, if the
potential is sufficiently flat near $\phi_{\rm top}$, stochastic
evolution can bring the field to the top, again with a rate given by
the HM bounce~\cite{Starobinsky:1986fx,Goncharov:1987ir, Linde:1991sk}

The situation with CDL bounces with endpoints near a potential 
maximum is similar.  If the potential maximum does not bound
the false vacuum potential well, then the CDL bounce is not
relevant for the decay of the false vacuum.  If, instead, the 
potential maximum is adjacent to the false vacuum well, the 
bounce, even if it has negligible overlap with the $|{\rm FV}\rangle$,
will overlap with thermally excited false vacuum states and will 
contribute to $\Gamma$; it can be understood as representing
thermally assisted tunneling.

Although a distant HM bounce may not be a reliable guide to the transition rate
out of the false vacuum, this does not mean that it has no physical
meaning; elegant answers in physics seldom lack a question.  Indeed,
the discussion above makes its significance clear.  The periodic path
integral with no constraints on configurations yields the partition
function for the full theory, not just the false vacuum sector.  The
contribution of a HM bounce to this path integral corresponds to the
probablity of finding the corresponding homogeneous configuration
within this thermal system.  The maxima at $\phi_B$ and
$\phi_{\rm top}$ in Fig.~\ref{secondfig} give HM bounces with equal
actions because the system is equally likely to be in a homogeneous
configuration with either of these three values.  Given an ensemble of
many three-sphere systems, all connected to a thermal bath, equal
numbers will be found to be approximately homogeneous with the field
equal to $\phi_B$ or $\phi_{\rm top}$.  Equivalently, if
one such three-sphere is followed for a long time, it will spend equal
amounts of time in each of these three configurations.  (Similar
statements should apply in the gravitational context, although
problems associated with defining a measure, as well as the loss of
regions from the quasithermal de Sitter system via decay to black
holes or to anti-de Sitter spacetime, must be dealt with.)  Note,
however, that this says nothing about the relative likelihood to go
directly from any particular configuration to each of these maxima.

The arguments presented in this note have addressed a particularly
striking anomaly associated with the HM solution.   However, there
remain a number of other unusual features associated with tunneling
in curved spacetime including, among others, the fact that the
CDL bounce is not sensitive to the shape of the potential in a
finite region including the initial vacuum.  It is clear that a
fuller account of curved spacetime tunneling is needed.  A partial
step in that direction will be presented elsewhere~\cite{brown}.

It is a pleasure to thank Adam Brown, David Gross, Bum-Hoon Lee,
Kimyeong Lee, and Piljin Yi for helpful conversations.  I am grateful
to Puneet Batra, Matthew Kleban, and Andrei Linde for their comments
on an earlier version of this article.  This work was supported in
part by the US Department of Energy.


\begin{thebibliography}{99}

\bibitem{Coleman:1977py}
  S.~Coleman,
  Phys.\ Rev.\ D {\bf 15}, 2929 (1977)
  [Erratum-ibid.\ D {\bf 16}, 1248 (1977)].

\bibitem{Callan:1977pt}
  C.~G.~Callan and S.~Coleman,
  Phys.\ Rev.\ D {\bf 16}, 1762 (1977).

\bibitem{Coleman:1980aw}
  S.~Coleman and F.~De Luccia,
  Phys.\ Rev.\ D {\bf 21}, 3305 (1980).

\bibitem{Hawking:1981fz}
  S.~W.~Hawking and I.G.~Moss,
  Phys.\ Lett.\ B {\bf 110}, 35 (1982).

\bibitem{volumefoot} The appearance of the flat-space factor of
   $4\pi/3$ in Eq.~(\ref{thermalB}) requires some comment, since the
   three-volume of a de Sitter horizon volume at constant time
   depends on the choice of coordinates, and is only given by the
   first factor in Eq.~(\ref{thermalB}) for a flat spatial slicing.
   In static coordinates, which are perhaps the most natural choice,
   the volume is $\pi^2 H_{\rm fv}^{-3}$.  However, in these
   coordinates the energy density includes a position-dependent
   redshift factor.  When this is taken into account, one finds that
   in the correct expression for the total energy in the horizon
   volume the $\pi^2$ factor is replaced by the ``naive'' $4\pi/3$.

   

\bibitem{Jensen:1983ac}
  L.~G.~Jensen and P.J.~Steinhardt,
  Nucl.\ Phys.\ B {\bf 237}, 176 (1984).

\bibitem{Hackworth:2004xb}
  J.~C.~Hackworth and E.~J.~Weinberg,
  Phys.\ Rev.\ D {\bf 71}, 044014 (2005).

\bibitem{Langer:1969bc}
  J.~S.~Langer,
  Annals Phys.\  {\bf 54}, 258 (1969).

\bibitem{Linde:1981zj}
  A.~D.~Linde,
  Nucl.\ Phys.\ B {\bf 216}, 421 (1983)
  [Erratum-ibid.\ B {\bf 223}, 544 (1983)].

\bibitem{firstfoot}
  In order to have a well-defined temperature on this finite system,
  as well as to allow the energy in the scalar field to vary, it must
  be coupled to some heat bath.  This can done either by introducing a
  large number of other fields on the three-sphere or by coupling it
  to some external (and possibly infinite) three-manifold. Note that
  it is no accident that this finite temperature, finite volume system
  displays the same difficulties as the potential with the de Sitter
  local maximum, since the gravitational HM bounce refers to a system
  with a finite horizon volume and a nonzero de Sitter temperature.

\bibitem{Banks:1973ps}
  T.~Banks, C.~M.~Bender and T.~T.~Wu,
  Phys.\ Rev.\ D {\bf 8}, 3346 (1973).

\bibitem{Banks:1974ij}
  T.~Banks and C.~M.~Bender,
  Phys.\ Rev.\ D {\bf 8}, 3366 (1973).

\bibitem{Batra:2006rz}
  P.~Batra and M.~Kleban,
  arXiv:hep-th/0612083.

\bibitem{Starobinsky:1986fx}
  A.~A.~Starobinsky,
in H.J.~De Vega and N.~Sanchez (eds.), {\it Field Theory, 
Quantum Gravity and Strings, 107-126} (Springer, 1986).

\bibitem{Goncharov:1987ir}
  A.~S.~Goncharov, A.~D.~Linde and V.~F.~Mukhanov,
  Int.\ J.\ Mod.\ Phys.\  A {\bf 2}, 561 (1987).

\bibitem{Linde:1991sk}
  A.~D.~Linde,
  Nucl.\ Phys.\  B {\bf 372}, 421 (1992).




\bibitem{secondfoot} 
  In the high temperature limit, where the dominant configurations are
  constant in imaginary time, this is equivalent to finding the number of
  critical bubbles per unit volume in the false vacuum (given by a
  Boltzmann factor) and multiplying by the characteristic time
  $\omega^{-1}$ (obtained from a determinant in the prefactor) for
  such bubbles to start expanding.

\bibitem{brown}  A.~R.~Brown and E.~J.~Weinberg, in preparation.







\end{thebibliography}
\end{document}